# The preparation of Zr-deuteride and phase stability studies of the Zr-D system

T. Maimaitiyili[a], A. Steuwer[b], C. Bjerkén[a], J. Blomqvist[a], M. Hoelzel[c], J. C. Ion[a], O. Zanellato[d]

tuerdi.maimaitiyili@mah.se

[a] Materials Science and Applied Mathematics, Malmö University, Nordenskiöldsgatan 1, 20506 Malmö, Sweden

[b]Nelson Mandela Metropolitan University, Gardham Avenue, 6031 Port Elizabeth, South Africa

[c]Forschungsneutronenquelle Heinz-Maier-Leibnitz (FRM II), Technische Universität Muünchen, Lichtenbergstr. 1, D-85747 Garching, Germany

[d]PIMM, Ensam - Cnam - CNRS, 151 Boulevard de l'Hôpital, 75013 Paris, France

**Abstract:**

Deuteride phases in the zirconium-deuterium system in the temperature range 25-286°C have been studied *in-situ* by high resolution neutron diffraction. The study primarily focused on observations of $\delta \rightarrow \gamma$ transformation at 180°C, and the peritectoid reaction $\alpha+\delta \leftrightarrow \gamma$ at 255°C in commercial grade Zr powder that was deuterated to a deuterium/Zr ratio of one to one. A detailed description of the zirconium deuteride preparation route by high temperature gas loading is also described. The lattice parameters of $\alpha$-Zr, $\delta$-ZrD$_x$ and $\varepsilon$-ZrD$_x$ were determined by whole pattern crystal structure analysis, using Rietveld and Pawley refinements, and are in good agreement with values reported in the literature. The controversial $\gamma$-hydride phase was observed both *in-situ* and *ex-situ* in deuterated Zr powder after a heat treatment at 286°C and slow cooling.

**Keyword:** Zirconium hydride, phase transformation, neutron diffraction, hydrogen induced degradation, high temperature hydrogen loading, deuterium

## 1 Introduction

Because of their good mechanical properties at high temperature and high pressure, their excellent corrosion resistance at moderate temperatures and low thermal neutron absorption cross section, zirconium (Zr) alloys are frequently used as a structural material in the nuclear industry [1-4]. As illustrated in Figure 1a, Zr-alloys are primarily used for fuel rod cladding to hold the fuel pellets inside the reactor core. The Zr alloy fuel rods are surrounded by a liquid that acts as both a coolant and neutron moderator. Depending on the type of nuclear power plant, either light ($H_2O$) or heavy water ($D_2O$) is used as a moderator. In ambient conditions, Zr reacts with oxygen and forms a thin (of the order tens of nm) passive oxide layer, which makes the alloy more corrosion resistant. However, during reactor operation the environment inside the reactor core becomes harsher (250-350°C, 7-15MPa) and Zr-alloys undergo corrosion, which produces free hydrogen (H) / deuterium (D). Some of the H/D atoms released through corrosion are absorbed into the cladding material. Some absorbed hydrogen may fill various faults in an area containing defects [5,6] and may become nucleation sites for hydrides [7]. Some of the hydrogen might be absorbed by Zr and may lead to the precipitation of hydrides once the solubility limit is exceeded [6,8]. Since Zr-hydride phases have a relatively large volume compared with the metallic Zr phase, the formation of hydrides introduces internal stresses in the material [9,10]. The local stress redistribution and plastic deformation induced by the growth of hydrides may contribute to the degradation of the material [11] and delayed hydride cracking [12-15]. Thus, the precipitation and growth of hydrides is a potential issue during extended fuel burn-up, and when reactors are taken off-line and cooled to ambient temperature.

It is known that the presence and orientation of hydrides degrades the mechanical properties of the material, with the degree of degradation depending on the concentration, distribution and morphology of the hydrides [13,16]. The detrimental effect of hydrides reaches a maximum when they are oriented normal to the applied load (radial hydrides in the case of tubes under internal pressure) [13]. Furthermore, different hydride phases exist in needle or plate like morphology depending on hydrogen concentration and environment [17,18]. In service, hydrides with different crystallographic structure and morphology might transform between each other under the influence of the surrounding environment (e.g., temperature and loading) [17,19], and react to external applied stress differently [20,21]. Therefore, to extend the life span of Zr-alloys and avoid failure, it is necessary to identify the nature of Zr hydrides and their exact structure.

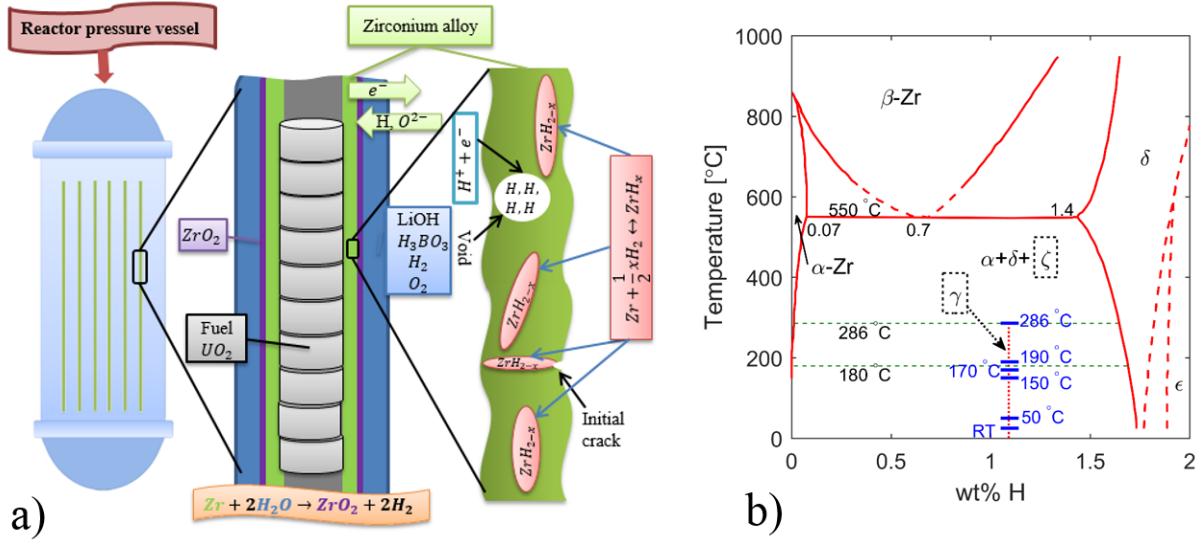

Figure 1: a) Illustration of nuclear fuel cladding and hydrogen related degradation; b) binary Zr-H phase diagram after [1]. Blue horizontal lines crossing the γ-ZrH phase line represent measurements carried out in this study. Both γ and ζ phases are poorly established and are indicated with dashed rectangle.

Since the binary phase diagram of the Zr-D system is principally the same as the Zr-H system (Figure 1b) [22,23] (D and H are chemically similar and there is no isotopic effect on phase boundaries), knowledge about Zr deuteride and Zr hydride can be used interchangeably for the purpose of phase transformation studies. However, deuterium has a higher coherent neutron scattering amplitude than hydrogen [22,24], which allows a better signal/noise ratio in neutron diffraction experiments. Therefore, deuterium is used in this work. Here D is used for hydrogen, and the term deuteride represents both hydride and deuteride.

At ambient pressure and temperature below 868°C, Zr has a hexagonal close-packed (HCP) structure known as α-Zr, which is of Mg structure type with $P6_3/mmc$ space group symmetry. The unit cell constants of α-Zr are $a$=3.2316Å and $c$=5.1475Å, and the $c/a$ ratio is 1.593, which is slightly below the ideal value 1.633 of perfect HCP but very close to the value of Ti $c/a$=1.587 [1-3,25-27]. There are two atoms in each primitive unit cell, one at 000 and the other at $\frac{1}{3}\frac{2}{3}\frac{1}{2}$ or at $\frac{2}{3}\frac{1}{3}\frac{1}{2}$ [28]. At higher temperatures (>868ºC), α-Zr allotropically transforms into the body-centred cubic (BCC) phase while the pressure is still low. This cubic phase is known as β-Zr phase, which is of W structure type with $Im\bar{3}m$ space group symmetry and a lattice constant $a$=3.6090 [1,25-27] (see Figure 1b). Under increased pressure while at room/low temperatures the HCP phase transforms into another hexagonal structure called ω-Zr phase ($c/a$=0.622) [26,27,29]. In contrast to α-Zr, the ω-Zr phase has three atoms instead of two [26,27,29].

The HCP structure of α-Zr consists of both tetrahedral and octahedral sites, and D is expected to diffuse and occupy these sites. However, according to experimental evidence, D tends to occupy tetrahedral sites [1-3,26,27]. According to most of the published literature [1-4,17,18,25-27], there exist at least four different deuteride phases (named δ-, ε-, γ- and ξ-deuteride) at ambient temperature under atmospheric pressure (Figure 1b and 2) depending on crystal structure, hydrogen concentration and quenching rate. The δ-deuteride (δ-ZrD$_x$) has a face centred cubic (FCC) structure with randomly positioned deuterium at eight tetrahedral sites with concentration ranges from $x$ = D/Zr = 1.5-1.7. The ε-deuteride (ε-ZrD$_x$) has a face centred tetragonal (FCT) structure with concentrations of $x$ = 1.74-2 (Figure 2). Both δ- and ε-deuteride are accepted as stable phases. Experimental evidence from the literature indicates that there is at least one more tetragonal phase existing in the Zr-D system with a Zr to D ratio of 1:1, which is denoted γ-deuteride (γ-ZrD) (Figure 2). However, controversy surrounds the stability and structure of the γ-ZrD phase. Some researchers suggest that γ-ZrD is the stable room temperature phase [30-33] with a δ to γ transition at 180°C [31] or transforms by a peritectoid reaction α+δ↔γ below ≈255°C [30], while others claim it is a metastable phase [13,34,35,36,37,38]. Additionally, several studies [1-4,22,39] propose that the γ-phase is a tetragonal phase with the $P4_2/n$ space group (Figure 2), but Kolesnikov et al. [40] reported that the structure of the γ-ZrD is orthorhombic, belonging to the space group *Cccm*. Many research findings report that the γ-ZrD phase occurs after rapid quenching, while the δ-ZrD$_x$ is often found after annealing [1,2,3,4]. In the meantime, the phase transformation order [34,36,41-45] between δ and ε together with the exact transition temperature [1,34,36,42-45] and precise deuterium concentration required [45-47] are also still debated. Furthermore, in addition to the aforementioned Zr-deuteride phases, another trigonal metastable Zr-deuteride phase with a stoichiometry of 0.25≤$x$≤0.5 has recently been reported, which belongs to the space group $P3m1$ ($a$=3.27Å, $c$= 10.83Å), and named ξ-deuteride [17]. ξ-deuteride is considered to be fully coherent with the α-Zr matrix and is speculated to be a precursor of δ and γ formation. Nevertheless, this new phase has not been confirmed or observed in this neutron diffraction or other *in-situ* neutron diffraction [22] and *in-situ* synchrotron X-ray diffraction studies [26,39]. Therefore, it is not considered further here.

As mentioned earlier, deuteride phases have a larger unit cell volume than the matrix, and the formation of deuterides introduces misfit strain that may contribute to cracking in the material. It has been reported that the volume expansions after phase transformations of α → γ and α → δ could be up to 12.3% and 17.2% per unit cell, respectively [9,48], and the plane expansion caused by ε-deuterides is about 7.1% [49]. Thus, knowledge of the precise structure, stability, formation and transformation behaviour of various $ZrD_x$ phases in the Zr-D system is essential for ensuring integrity of the Zr-based materials in power plant.

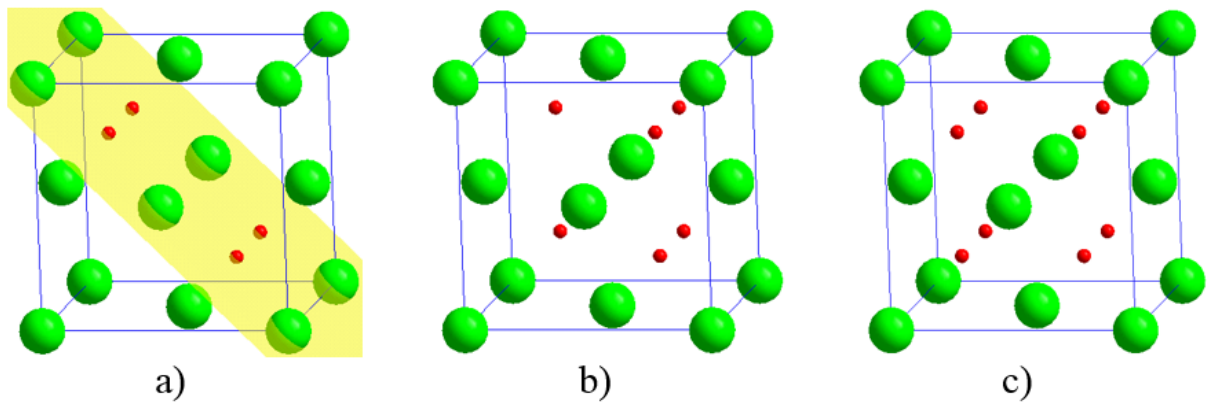

Figure 2: Crystal structure of (a) γ-, (b) δ- and (c) ε-Zr deuteride. Green spheres represent Zr atoms and red spheres represent deuterium atoms.

The discrepancy concerning $ZrD_x$ phase stability, crystal structure and transformation may arise because of: the high diffusivity of deuterium at low temperature (<600°C) [26,50]; close structural similarities between relevant phases [26,50]; non-stoichiometry (0<$x$≤2) of deuteride phases [1-4,17,25-27]; the strong influence of other impurity elements in the Zr-D system [1,4,27]; an unclear deuteride preparation route [1,4,26,27,30] and the sensitivity of experimental methods [39]. Therefore, to overcome experimental complications arising from the sample and instrument, high resolution *in-situ* experiments of high purity samples is essential. Here a preparation route of the ZrD through high temperature gas loading is described. Results from high-resolution, neutron diffraction experiments performed on the Structure POwder DIffractometer (SPODI) [51] instrument at Forschungsneutronenquelle Heinz Maier-Leibnitz (FRM II), Garching, Germany are reported. The objectives of the work were to investigate the γ-ZrD preparation route; the existence of a transition between δ-$ZrD_x$ and γ-ZrD; the transition temperature between δ-$ZrD_x$ and γ-ZrD; and the stability of δ-$ZrD_x$ and the ε-$ZrD_x$.

## 2. Material and method

### 2.1 Materials and deuteration

Both neutron and X-ray diffraction techniques enable absolute atomic spacings in crystals to be determined; techniques that are extensively used for establishing crystallographic properties and defining atomic structures. Neutrons possess unique scattering amplitudes for each element isotope [24]. The scattering length of D, for example, is significantly larger than for H [24]. Hence, by using deuterated Zr-powder it is possible to obtain a large signal to noise ratio even at low concentrations of D with a high resolution neutron diffraction facility such as SPODI at FRMII.

Commercial grade Zr powder (Table 1) was deuterium-charged by using a high temperature gas diffusion setup shown in Figure 3, with several pressure increases and drop steps as shown in Table 2. Firstly, commercial grade pure zirconium powder (99.2% purity) with a maximum particle size of ≈45μm was obtained from Goodfellow Ltd., Huntingdon, England (ZR006015), and placed in a glass container (Figure 3). To prevent contamination or oxidation all powder handling was performed inside a glove box in an argon environment. To ensure optimal D diffusion and heat transfer, the powder was spread evenly inside the container. Subsequently, the complete loading system including the sample was baked under vacuum for three days at 300°C until the system pressure decreased to $5\times10^{-14}$ MPa. Thereafter, the sample chamber was filled with a given amount of D. When the chamber pressure had attained a low level (D absorbed into the powder), the chamber was refilled. This cycle was repeated until the desired D and Zr (D/Zr) ratio 1:1 was achieved. Here it should be noted that the loading chamber temperature was kept at 300°C during whole process. The nominal composition of the sample is shown in Table 1, and details of each Zr-powder deuteration steps are tabulated in Table 2.

Table 1. Nominal composition of the Zr-powder specimens studied in this study.

| wt% | Impurity elements [ppm by weight] | | | | | | |
|---|---|---|---|---|---|---|---|
| Zr | C | Hf | Fe | Cr | N | O | H |
| 99.2 | 250 | 2500 | 200 | 200 | 100 | 1000 | 10 |

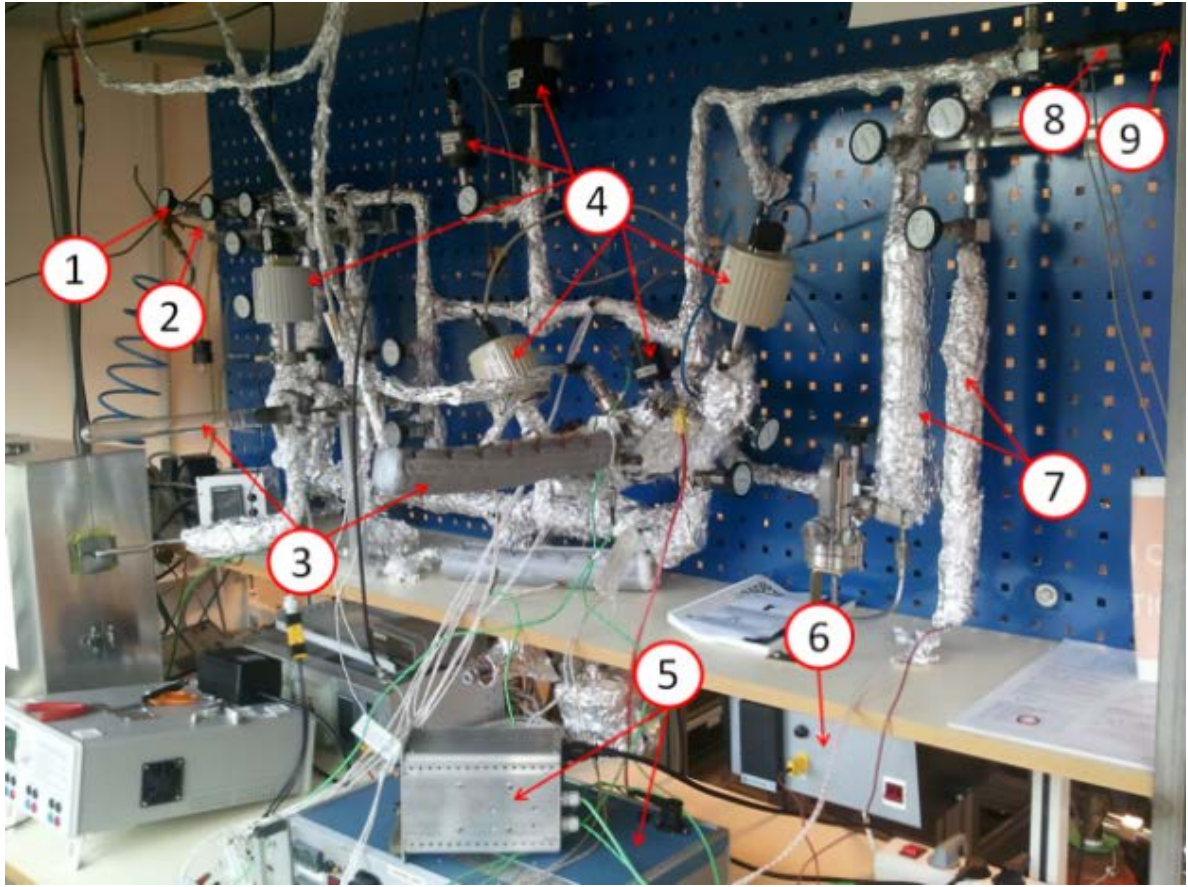

Figure 3: Zirconium powder loading setup. (1) External $D_2$; (2) $D_2$ loaded metal powder (not connected); (3) UHV sample chamber; (4) Pressure gauges; (5) Temperature controllers; (6) Metal powder heating unit; (7) $H_2$ loaded metal powder; (8) $H_2$ purifier; (9) External $H_2$ source. (7)-(8) are not used.

Table 2: The Zr powder loading steps.

| Deuterium filling steps | Pressure before absorption [kPa] | Pressure after absorption [kPa] |
|---|---|---|
| Step 1 | 189.6 | 6.9 |
| Step 2 | 190.2 | 12.0 |
| Step 3 | 190.2 | 9.6 |
| Step 4 | 193.6 | 19.5 |
| Step 5 | 210.0 | 1.1 |

The deuterium content in the sample was determined by both mass and pressure change, and found to be 0.969(1) and 1.0(3) by atomic ratio, respectively. Mass measurement was repeated three times. For simplicity, here we do not address questions related to the kinetics of hydrogen desorption/absorption from zirconium hydride. Information about kinetics and similar gas loading setups can be found in [52] and [53].

## 2.2 Experimental design and data analysis

### 2.2.1 *In-situ* studies

The thermal, high-resolution powder diffractometer SPODI at FRMII was used for the investigation. This specific neutron beam line was selected because of its high diffraction resolution with low signal-to-noise ratio, which is ideal for Rietveld analysis [54].

The SPODI was equipped with a germanium (551) focusing monochromator with mosaicity 20' horizontally and 10' vertically. The detector array consist of 80 $^3$He position sensitive detectors with fixed Soller collimator of 10' divergence, which are arranged radially around the sample position in the centre. The active height of the detectors is 300 mm and the separation angle between each detector is 2º. As each detector covers 2° (160° / 80), the multidetector of SPODI spans an angular range of 2θ = 160°, and the finest resolution can be achieved with a step width $\Delta(2\theta)$=0.05º [51].

In order to ensure quality of the data refinement, in 40 steps (160°/[80 $^3$He detectors × $\Delta(2\theta)$]=40) of angular shift with several overlaps and 300 000 counts for each measurement were used. The neutron wavelength was calibrated with a Si standard to λ=1.54819(91)Å and was fixed during the entire experiment. Since a Soller collimator is present in the setup, the diffraction data were corrected for beam divergence after the measurement. To investigate the stability of various deuteride phases, measurement started from room temperature, and the data were collected continuously as a function of temperature and time. Once the system temperature reached 286°C, conditions were kept constant for 5 h until no further phase transformation was observed. The detail temperature and data acquisition times are shown in Table 3. The system was then allowed to cool to room temperature at the same rate. Both heating and cooling rates were ~10ºC/min. In order to investigate possible γ→δ transition temperatures, additional measurements were recorded around 180 and 250°C.

Table 3: *In-situ* neutron diffraction data acquisition time and corresponding temperature.

| **Temperature [°C]** | 25 | 50 | 150 | 170 | 190 | 286 | 286 | 150 | 25 |
|---|---|---|---|---|---|---|---|---|---|
| **Time [hours]** | 1 | 1 | 3 | 3 | 3 | 3 | 3 | 9 | 3 |

### 2.2.2 *Ex-situ* studies

Quartz capillaries were filled with deuterated Zr-powder inside a glove box under an Ar atmosphere and placed into a quartz Schlenk tube. Dynamic pumping was used to evacuate trapped Ar. Subsequently, the packed powder was heated inside a furnace from room temperature (28ºC) to 286ºC at a heating rate of 10ºC /min. The sample was then held at 286ºC for three hours and allowed to cool to room temperature in the furnace at the same rate. Finally, the heat-treated powder after all thermal cycles was placed into a glove box again and packed into capillaries, which thereafter were sealed with wax for future measurements.

Conventional laboratory X-ray diffraction experiments were performed using the STOE STADI MP diffractometer in transition mode using Cu *Kα*$_1$ radiation (λ=1.5406Å), a Ge monochromator and a Mythen detector. The data were collected in the 2θ range 2-90º for 30 minutes with an acquisition rate of 30s/step, where a step corresponded to 0.9 in 2θ.

### 2.2.3 Profile analysis and refinement

In order to verify the existence of various phases, the whole pattern structure analysis software packages Topas-Academic [55] and GSAS [56] were employed to perform Rietveld analysis.

The instrumental function and wavelength of the neutron beam calibrated with a standard Si sample was collected with the same setup prior to measurements. The peak shape was described by a pseudo-Voigt function [55]. To describe the influence of beam divergence on reflection profiles, a simple axial model [55] is also included in the refinement. The background of the diffraction pattern was fitted using a Chebyshev function [55] with five coefficients and the zero shift error calibrated from Si was refined only for first diffraction patterns and then fixed. As the material in question is of high purity and fine grain powder, the preferred orientation/texture was not considered. The scale factor, lattice parameters and isotropic atomic displacement parameters of the phases present in the system were refined. Atomic positions of none of the phases were refined. The occupancy of the deuteride atoms was refined only once for the first diffraction patterns at the last stage of the refinement and then fixed. Here we should emphasize that in this work we did not intend to solve any structure and determine absolute lattice parameters. Instead the focus is on phase identification and relative changes.

As it is commonly believed that the Zr deuteride system consists of at least four phases ($\alpha$, $\delta$, $\varepsilon$ and $\gamma$), the data analysis commenced with this assumption. The metallic Zr phase $\alpha$ and the deuteride phase $\delta$ were fitted with space groups $P6_3/mmc$ (No.194), with cell parameters $a$=3.2317Å, $c$=5.1476Å, and $Fm$-$3m$ (No.225), with cell parameters $a$=4.7803Å, respectively [1]. Because of the uncertainty about the structure of the $\gamma$-deuteride phase, the Rietveld refinement for the $\gamma$-phase was carried out by using two different structures: the first structure of $\gamma$ was treated as face centred tetragonal (FCT) with space group $P4_2/n$ (No.86) and cell parameters $a$=4.5957Å, $c$=4.9686Å [1]. The second structure was considered orthorhombic with the space group symmetry $Cccm$ (No.66) and cell parameters $a$=4.549Å, $b$=4.618Å, $c$=4.965Å [40]. The fitting results of orthorhombic assumption were not satisfactory. Therefore, it is not discussed further here. A relevant discussion regarding the observation of the orthorhombic structure can be found in [2]. Basic unit cell parameters of all these phases used as reference in data our refinement procedure and corresponding refined values are listed in Table 4.

Despite careful powder handling and high purity of the Zr powder, visible traces of Zr oxide and second phase particles (SPP) were observed in all diffraction patterns, which can be observed in Figure 4, 5, 6 and 8. Therefore, in all refinements both oxide and SPP were refined together with other phases. To avoid confusion, their $hkl$ indices are only shown in Figure 5 and 6.

Table 4: Known and fitted unit cell lattice parameters of various Zr, Zr-deuteride and other secondary phases at room temperatures from *in-situ* neutron diffraction data.

| Phase | Structure | Space group | $a$ [Å] | $b$ [Å] | $c$ [Å] | Reference |
|---|---|---|---|---|---|---|
| $\alpha$(Zr) | HCP | $P6_3/mmc$ | 3.2317 | 3.2317 | 5.1476 | [1] |
| | | | 3.2344(4) | 3.2344(4) | 5.1507(5) | *Present studies |

| | | | 3.2349(3) | 3.2349(3) | 5.1512(1) | Present studies |
|---|---|---|---|---|---|---|
| $\delta(ZrD_{1.66})$ | FCC | $Fm\bar{3}m$ | 4.7803 | 4.7803 | 4.7803 | [1] |
| | | | 4.7795(4) | 4.7795(4) | 4.7795(4) | *Present studies |
| | | | 4.7735(1) | 4.7735(1) | 4.7735(1) | Present studies |
| $\gamma(ZrD)$ | FCT | $P4_2/n$ | 4.5957 | 4.5957 | 4.9686 | [1] |
| | | | 4.5930(4) | 4.5930(4) | 4.9564(4) | Present studies |
| | Orthorhombic | $Cccm$ | 4.549(1) | 4.618(1) | 4.965(1) | [40] |
| $\varepsilon(ZrD_2)$ | FCT | $I4/mmm$ | 4.9757 | 4.9757 | 4.4510 | [1] |
| | | | 4.9652(9) | 4.9652(9) | 4.4567(2) | *Present studies |
| #$ZrO_2$ | Monoclinic | $P12_1/c1$ | 5.168 | 5.232 | 5.341 | [57] |
| | | | 5.1639(3) | 5.2692(2) | 5.3374(5) | Present studies |
| $ZrFeCr_2$ | HCP | $P6_3/mmc$ | 5.012 | 5.012 | 8.223 | [58] |
| | | | 4.9773(1) | 4.9773(1) | 8.2371(3) | Present studies |

Note: * At room temperature before heat treatment; #Unit cell angles of $ZrO_2$ are $\alpha=90°$, $\beta=99(2)°$ and $\gamma=90°$.

## 3. Result and discussion

The series of neutron powder diffraction patterns of the deuterated Zr powder collected at various temperatures during *in-situ* heating experiment are shown in Figure 4. The heat treatment and the temperatures are specified on left of the figure. As noted, the first diffraction pattern at the bottom corresponds to untreated, as received data at room temperature; the top most are the data acquired at room temperature after all heating cycles. Non-overlapping/separated and clear reflections of $\gamma$ and $\varepsilon$ Zr deuterides are also labelled using Miller indices on top of each corresponding peak position. Estimated peak positions of all these phases present in the system are marked with colour coded vertical bars on the bottom of the diffractograms. Four clearly separated *γ* phase reflections are also marked with vertical dashed lines with the same colour with *γ* phase Bragg peak position markers.

Similar diffractograms are plotted with the same colours in Figure 4. As illustrated in Figure 4, the complete thermal cycle can be divided into four stages. The first is the pre-phase transformation region from room temperature to 190ºC during which all the diffraction patterns (shown in black) are almost identical to the as-received samples. The second stage is the initiation of the *ε→δ* transformation at 286ºC. The third stage represents thermal equilibrium at 286ºC after a dwell time of five hours. The final stage includes *γ* formation and growth at 150 and 25ºC during cooling. From this figure it can be clearly seen that all the peak positions and shapes of the diffraction patterns in the temperature interval 25-190°C are identical, but when the system temperature reaches a temperature of 286°C some of the existing reflections disappear and new reflections appear. This implies that there are no phase transformations between 25 and 190°C; instead phase transformations take place around 286°C. From Figure 4 it can also be seen that none of the diffraction patterns obtained at 286, 150 and 25°C during cooling reveal any *ε* phase reflections.

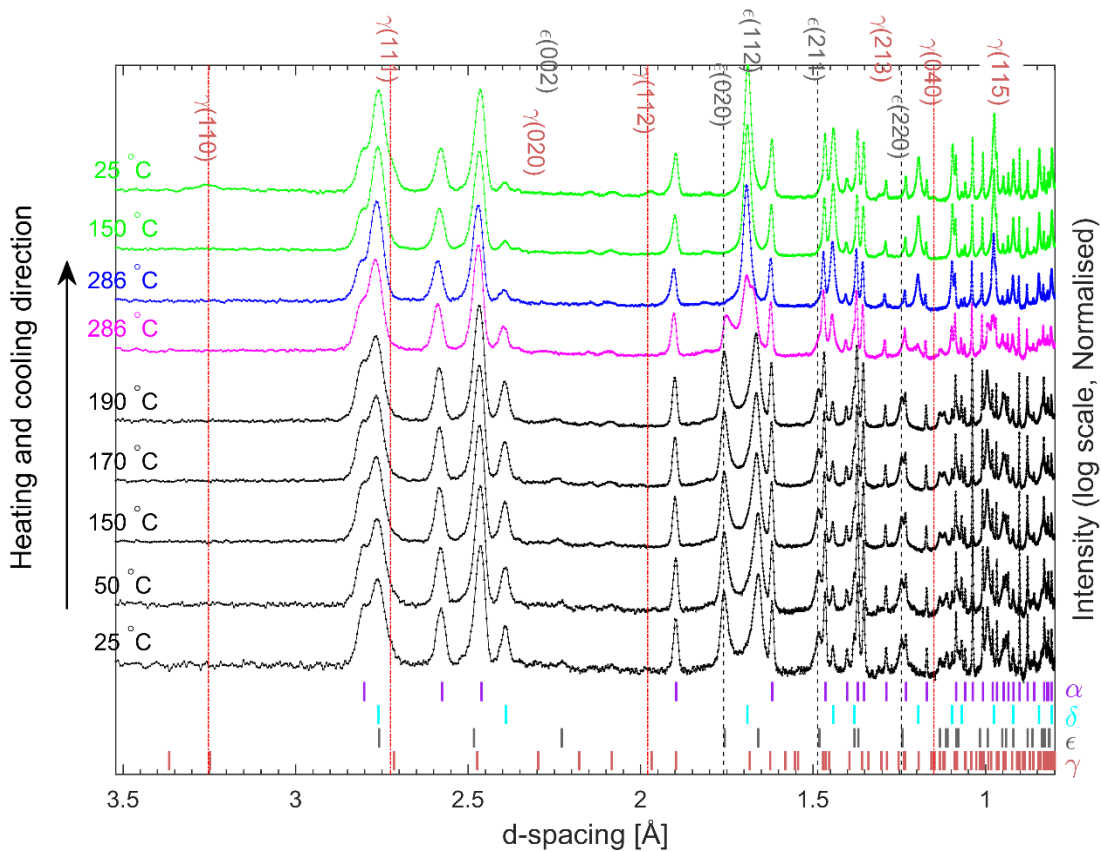

Figure 4: Neutron diffraction patterns of deuterated Zr powder at various temperatures during *in-situ* heating. The colour coded, small vertical tick marks represent the *hkl* peak positions of labelled phases in the same colour. For clarity of some of the diffraction peaks, the reader is referred to the web version of this article.

The results of Rietveld refinements of the data acquired at room temperature before any heat treatment (as received) and after heat treatment are shown in Figure 5 and 6, respectively. The refined structural parameters of various phases present in the sample are tabulated in Table 4. In both Figures 5 and 6, the blue diffraction pattern is the observed data and the red pattern represents the calculated pattern. The quality of the refinement can be seen by the difference plot in black presented in the middle and the weighted R ($R_{wp}$) values given in the upper left of the figure. From the difference plot and the $R_{wp}$ (weighed residual factor), it can be seen that the fitting results are satisfactory. According to the Rietveld analysis of neutron powder diffraction data from the as-received sample, the main constituents were the $\delta$- and $\varepsilon$-ZrD$_x$ phases together with the $\alpha$-Zr phase. In addition, trace amounts of ZrO$_2$ and ZrFeCr$_2$ were also observed at room temperature prior to heat treatment, but there was no sign of $\gamma$-ZrD in the system at this stage.

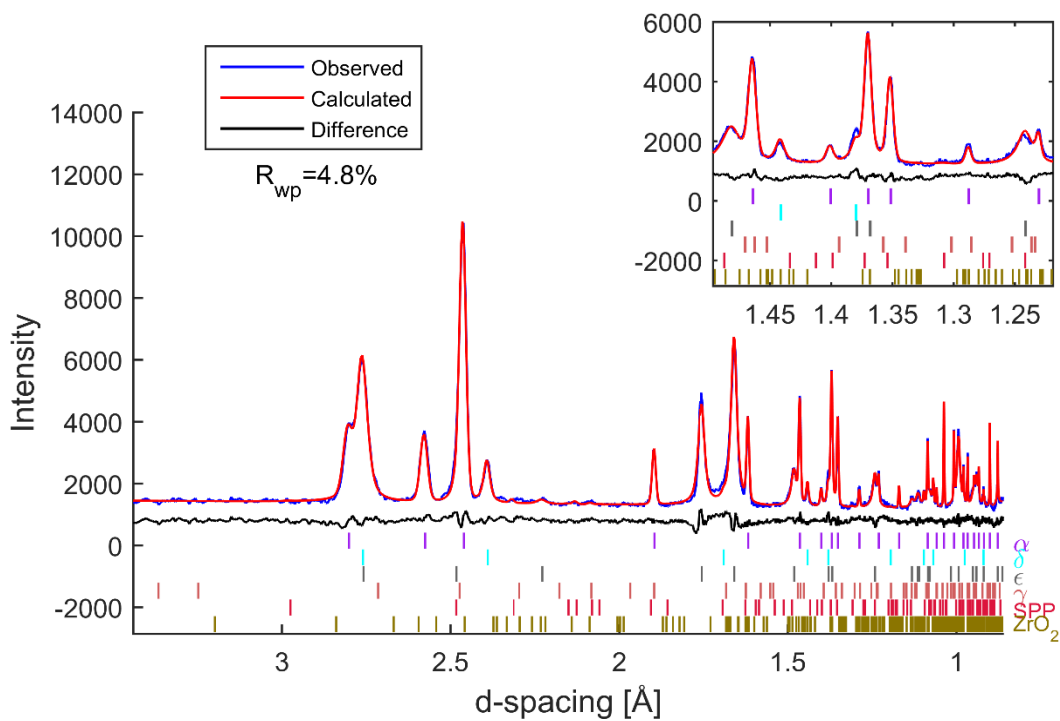

Figure 5: Rietveld refinement of powder neutron diffraction data of deuterated Zr powder at room temperature in as received state prior to heat treatment. Observed, calculated and Bragg

reflection positions are shown by blue, red solid line and colour coded bars, respectively. The difference plot is shown in the middle as a solid black line. The inset shows the expanded plot from $d$-spacing = 1.22 to 1.50Å. The goodness of fit (GOF) is 1.820.

Figure 6 shows neutron diffraction plots together with the profiles based on the Rietveld fits of various phases collected at room temperature after heat treatment. The figure shows that the system contains about 9.55wt% of $\gamma$-ZrD phase. The structure matches that of the crystal structure of FCT with space group P4$_2$/n. During refinement, the occupancies of the deuterium and Zr atoms were fixed to unity. The cell parameters of the $\gamma$ phase were calculated to be $a$=4.5930(4)Å, $c$=4.9564(4)Å.

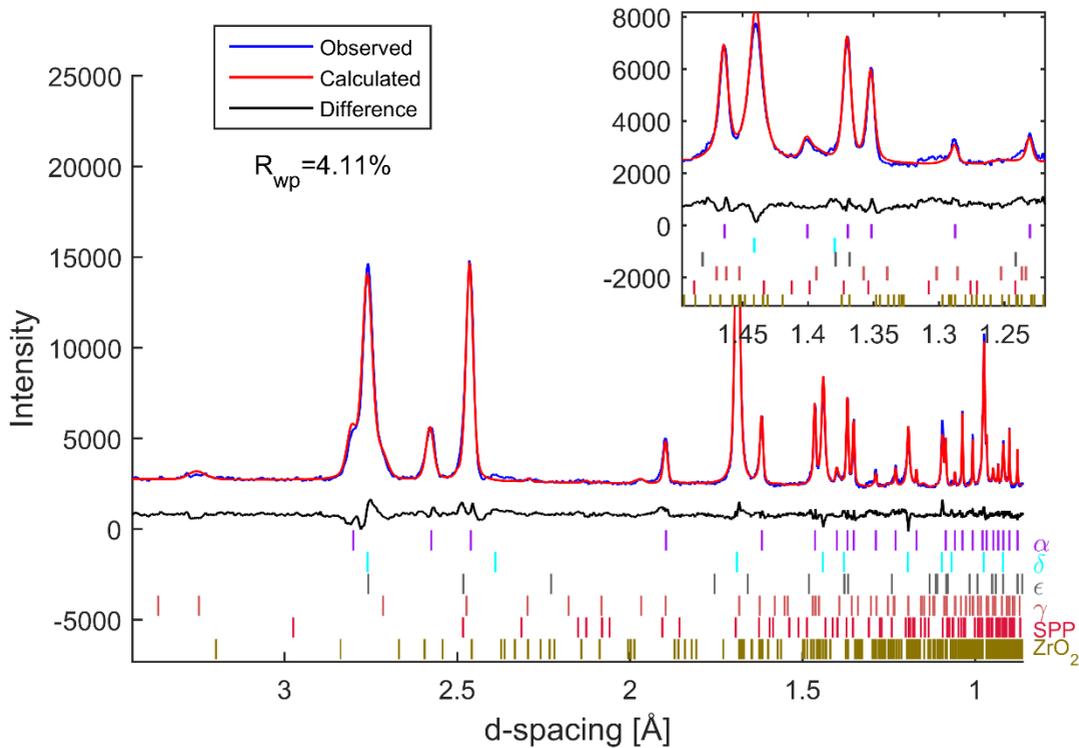

Figure 6: Rietveld refinement of powder neutron diffraction data of deuterated Zr powder at room temperature after all heat treatments. Observed, calculated and Bragg reflection positions are shown by blue and red solid lines, and colour coded bars, respectively. The difference plot is shown in the middle with a solid black line. The inset shows the expanded plot for $d$-spacing=1.22 to 1.50Å. The goodness of fit (GOF) is 2.1.

Careful inspection of Figure 5 and 6 reveals that at room temperature after annealing there are clear shoulders next to some of the $\alpha$-Zr reflections (e.g., at $d$-spacing=1.30, 1.23, 1.92). Any attempt to fit these shoulders with all currently reported deuteride phases including $\zeta$-ZrH$_{0.25-0.5}$ [17], common Zr oxides and Zr nitrides did not yield satisfactory results. To the best of our knowledge these shoulders do not represent deuterides or second phase particles (SPP).

Quantitative estimation of $\alpha$-, $\delta$-, $\gamma$- and other minority phases were made using the mass volume relationship [55]:

$$W_p = S_p(ZMV)_p \Big/ \sum_{i=1}^{n} S_i(ZMV)_i$$

where $W$=relative weight fraction of phase $p$; n=number of phases; $S$=scale factor; $Z$=atomic mass; $M$=number of formula units per cell; $V$=unit-cell volume. Figure 7 summarizes the phase weight fraction variation of each phase during complete heat treatment in the Zr powder that was obtained after the Rietveld refinement. Weight fractions are colour coded to maintain consistency with previous figures. As the total phase fractions of SPP and $ZrO_2$ were small (<2 wt%) and they are not of primary interest of this study, they are not presented.

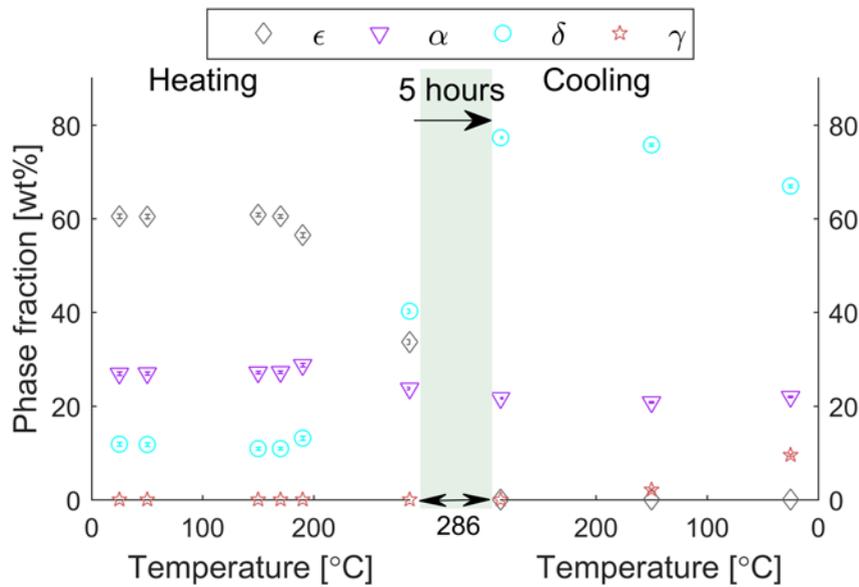

Figure 7: Phase fraction variations of various Zr and Zr deuteride phases in the system. Error bars are also given in the figure, which are smaller than the markers.

In order to confirm the existence and as well as the possible reproducibility of $\gamma$ Zr deuteride observed after thermal treatment in *in-situ* neutron diffraction studies, the diffraction patterns collected at room temperature from samples prepared for *ex-situ* studies and collected with an X-ray diffractometer are compared with one external *in-situ* data set reported in [50] which showed clear evidence of $\gamma$-ZrD. As shown in Figure 8, the peak position of $\gamma$ in all three diffractograms agreed well and the Rietveld analysis also showed a comparable amount of $\gamma$ in the *ex-situ* data. From Figures 7 and 8, it is apparent that there is a considerable amount of *γ* deuteride phase in the powders after heat treatment. This implies that the deuteride preparation route that has been developed is functional.

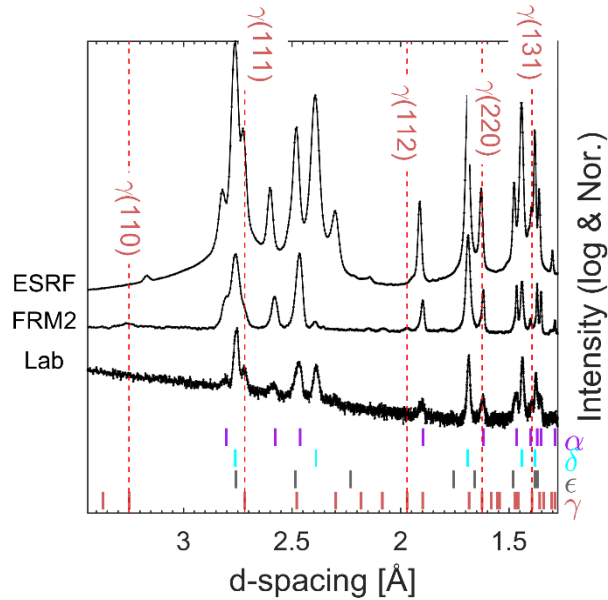

Figure 8: Comparison of diffraction patterns collected at room temperature at different facilities from samples heat-treated slightly differently. The pattern labelled 'ESRF' is the external synchrotron X-ray data from [50]; the pattern 'FRM2' is the neutron data from FRM II and 'Lab' is from a conventional X-ray diffractometer.

Despite the reported γ formation composition (D/Zr ratio 1:1) [30,59], no 100% γ Zr deuteride was obtained during either *in-situ* and *ex-situ* studies. In fact, both measurements yielded only about 10% γ, disregarding different cooling rates.

The concentration of alloying elements in the Zr matrix is limited, and therefore the observed volume fraction of the SPP in the Zr is also very small (0.1(9)-0.7(5)%). Because of the diffraction background contribution by the environment and diffuse scattering from the matrix, it is challenging to perform any quantitative studies on SPP in Zr with confidence. However, as shown in all the diffractograms presented here, with high resolution neutron powder diffraction it is possible to obtain high quality data and perform quantitative studies on such limited quantity phases. From the analysis it is noted that the quantity of the SPP did not change significantly on heating and cooling. The $ZrO_2$, however, slightly increased during heating but did not vary during cooling. To make quantitative determinations of SPP and $ZrO_2$ and investigate their role on deuteride phase stability and transformation in the Zr-D system, we suggest including standard reference material in similar future studies.

From Figure 4 it is difficult to see if there is a clear phase interchange between $\delta$- and $\varepsilon$-ZrD$_x$ phases below 286°C. However, according to Figure 7, it appears that some transformation started to take place at 190°C. When the temperature reached 286°C, the intensity and phase fraction of the $\varepsilon$-phase decreased significantly while the intensity and phase fraction of the $\delta$-phase increased, which is apparent in both Figures 4 and 7. Since there are no diffraction patterns collected between 190°C and 286°C, it is very hard to confirm that the transformation between $\delta$- and $\varepsilon$-ZrD$_x$ actually started at 190°C, and there is a continuous transformation until 286°C. According to *in-situ* hydrogenation studies carried out on synchrotron X-rays [26,41,50] and as well as other literature [1,44,60,61], it is safe to assume that there is a smooth/gradual change between $\delta$- and $\varepsilon$-ZrD$_x$ in the temperature range 190-286°C. The phase percentage variations of $\delta$- and $\varepsilon$-ZrD$_x$ in Figure 7 also support such gradual transformation. Here we would like to point out that, even though there are no diffraction patterns collected between 190°C and 286°C, the real time data was inspected frequently and during these inspections we did not observe evidence of any new phases or dramatic changes in diffraction patterns. As the phase transformation become more evident once the temperature reaches 286°C, the system temperature was maintained at 286°C for five hours to allow the phase transformation to be completed. After five hours, when no further changes were observed, the system temperature was decreased to observe whether the transformed $\varepsilon$-phase transformed back. This reversal was not observed in this study. As shown in Figure 7, starting at 11.8(9)wt% at room temperature, the amount of $\delta$-phase increased as $\varepsilon$-phase decreased. At 286ºC after five hours of baking all of the $\varepsilon$-phase transformed into $\delta$-phase (77.3(2)wt%). The amount of the α-Zr phase during the whole experiment was approximately consistent before (27.2(5)wt%) and after (22.0(3)wt%) the $\varepsilon \rightarrow \delta$ transformation, with only a small loss occurring during this hydride transformation.

Variation of the $\varepsilon$-ZrD$x$ lattice parameters with respect to temperature and phase transformation showed close similarity to reported observation in [41] (Figure 9b), and agreed with data presented in Figure 7. As presented in Figures 4 and 7, the transformation from $\varepsilon$ to $\delta$ started from a mixture of three phases ($\alpha+\delta+\varepsilon$) and was completed after several hours at 286ºC. However, the deuterides that transformed to $\delta$-ZrD$_x$ did not transformed back to $\varepsilon$-ZrD$_x$ after the system temperature returned to room temperature. This all together indicated that the sample was not homogeneous and not in the equilibrium state in the beginning, but after transformation the deuterium concentration homogenized and reached equilibrium.

The refined lattice parameters of $\alpha$-Zr, $\delta$-ZrD$_x$ and $\varepsilon$-ZrD$_x$ are plotted in Figure 9. It is seen that the lattice parameters $a$ and $c$ of the $\alpha$-Zr phase essentially maintain a linear relation with temperature during heating prior to phase transformation at 190°C. The lattice parameters of the $\alpha$-Zr, $\delta$-ZrD$_x$ and $\varepsilon$-ZrD$_x$ obtained in this study before and after heat treatment are in good agreement with reported values [1] (Table 4). From Figure 9, it can be seen that there is a hysteresis in the lattice parameters of $\alpha$-Zr and $\delta$-ZrD$_x$ before and after annealing. This may be caused by an apparent D solubility limit difference between heating and cooling in the Zr [15] or it may simply be caused by diffusional lag.

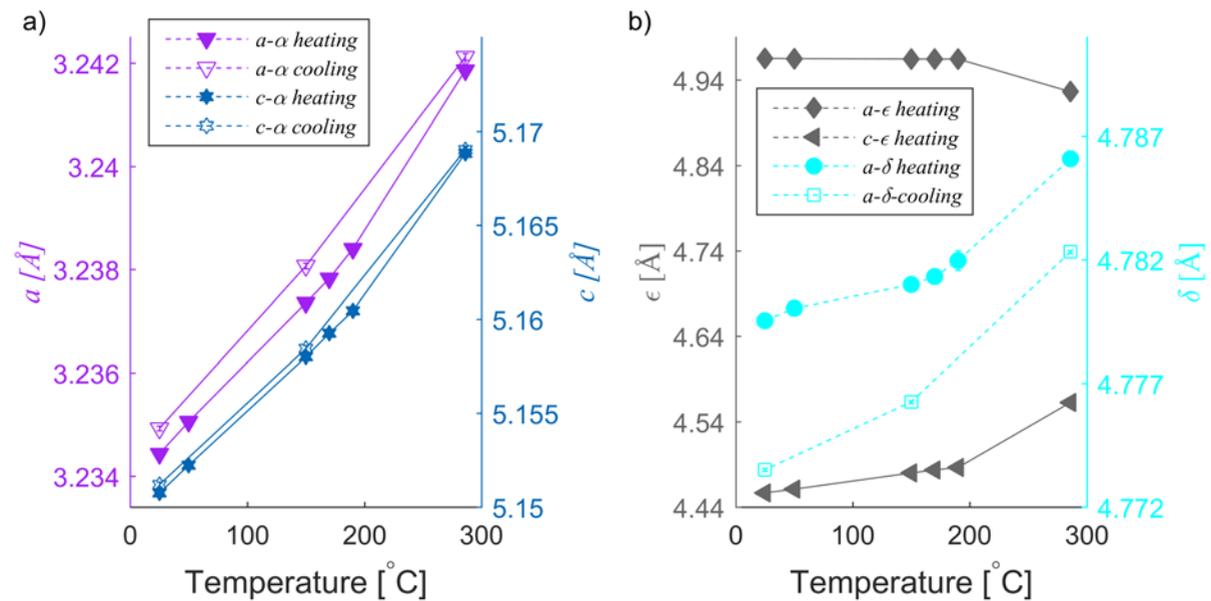

Figure 9: Variation of a) $\alpha$-Zr and b) $\delta$-ZrD$_{1.66}$ and $\varepsilon$-ZrD$_2$ lattice parameters as a function of temperature. In subplots, filled markers represent heating, and unfilled markers represent cooling. Lines connecting data points are used to highlight the linear relationships in (a) and (b). The error bars are also plotted, but are smaller than the markers.

Mishra et. al [30] has reported that there is peritectoid reaction at about 255 °C between the $\alpha$ and $\delta$ phases which gives the $\gamma$ phase ($\alpha+\delta\leftrightarrow\gamma$ transformation), and the $\gamma$ phase is a stable phase of composition ZrD below 255°C. However, none of the diffraction patterns collected before and at 286°C showed any sign of $\gamma$-ZrD (Figures 4 and 7). Nonetheless, neutron powder diffraction patterns collected at 150°C and room temperature, after heat treatment at 286°C showed the presence of *γ*-ZrD (Figure 4). The Rietveld analyses also supported the existence of *γ*-ZrD (Figure 7). The exact reason why *γ*-ZrD is not observed before heat treatment is not clear. We do not know whether this is related to the homogeneity of the hydrogen concentration in the system or the presence of the *ε*-ZrD$_x$, which prohibits its formation.

According to Figure 7, the formation and growth of *γ*-ZrD is compensated for by dissolution of *δ*-Zr deuterides. This observation supports the existing observation of slow *γ* Zr deuteride formation by *δ*→*γ* transformation at 180°C proposed by Root et al. [31] and a recent report [22,39]. In line with *in-situ* neutron diffraction studies, the laboratory X-ray data also did not show the presence of the *γ*-ZrD in untreated deuterated powder at room temperature, but it did reveal strong *γ*-ZrD presence at room temperature after similar heat treatment used in *in-situ* neutron diffraction studies.

The exact cause of the *δ*→*γ* phase transformation temperature mismatch in this and other literature may be related to the purity of the material or the hydrogen concentration as discussed in [1,4]. Considering the presence of SPP and slight variations in oxides in the system, it may be reasonable to assume that a correlation exists between hydride and minority phases. It may be that these SPP provide nucleation sites for *γ* Zr deuteride or act as catalysts facilitating its formation. Moreover, both [1] and [60] have reported that the size of the two-phase (*δ*+*ε*) region in the Zr-H system might be influenced by the oxygen as well as other impurities in the Zr. In addition, it also been reported that impurities play a decisive role in Zr deuteride stability [1,4,60]. Therefore, it is very difficult to establish a firm relation about the exact formation mechanism of the *γ* Zr-deuteride, which was beyond the scope of this work.

In a nuclear reactor environment, *δ*-ZrD$_x$ is the most commonly seen deuteride and is considered to be most detrimental to the integrity of the cladding [4,45,62,63]. For that reason, in the past insufficient attention had paid to investigating low and high D concentration phases, such as *γ*-ZrD and *ε*-ZrD$_x$ respectively. However, this work highlighted that the *ε*-ZrD$_x$ may transform into *δ*-ZrD$_x$, then *δ*-ZrD$_x$ can slowly transform into the *γ*-ZrD after heat treatment despite slow cooling rates. In addition, the work also showed the existence of the *γ*-ZrD at low temperature (<190°C) for an extended period of time in the system. Therefore, the presence of γ-ZrD and ε-ZrD$_x$ may also be important to the formation and evolution of the δ-ZrD$_x$. According to Steuwer et al.[2], under the influence of stress, the *δ*-ZrD$_x$ may transform into *γ*-ZrD via ordering of the D atoms. Since the *γ*-ZrD was reported as needle like in shape [17] and deuterides have a tendency to form perpendicular to applied stress [19], the integrity of the cladding not only depends on δ-ZrD$_x$ but requires a good understanding of γ-ZrD.

## 4. Conclusions

Observations of *δ→γ* transformation at 180 and 255°C have been studied both *in-situ* and *ex-situ*. The present study has not only demonstrated the use of neutron diffraction for *in-situ* measurement of Zr deuteride phase transformation and identification, but also gives a detailed description of a deuteride preparation route used to generate powder with an atomic ratio of 1:1 of D:Zr.

γ-ZrD is observed at room temperature after slow cooling (10°C/min) from a 5 hour heat treatment at 286°C. No formation of *γ* phase through a peritectoid reaction at 255°C was observed during the heating or cooling cycles of the *in-situ* measurements. Instead, the *δ→γ* deuteride transformation was observed. It is further suggested that the formation of *γ* phase is a slow process as reported in literature [22,26,31,39]. Furthermore, the lattice parameters of *α*-Zr, *δ*-ZrD$_x$ and *ε*-ZrD$_x$ are found to be in good agreement with values reported in the literature [1]. The crystal structures of *δ*-ZrD$_x$ show isotropic thermal expansion, while the deuterium richer *ε*-ZrD$_x$ shows anisotropic thermal expansion. The γ-ZrD repeatedly synthesized using *ex-situ* heat treatment and *ex-situ* laboratory X-ray diffraction results substantiate the observations made during *in-situ* neutron diffraction.

## 5. Acknowledgments

The FRMII is gratefully acknowledged for the provision of beam time. We are thankful to the Swedish Research Foundation (VR 2008-3844) for financial support. Professor S. Lidin at the Division of Polymer & Materials Chemistry at Lund University is gratefully acknowledged for providing access to his laboratory X-ray diffractometer and valuable discussions.